\documentclass{ws-procs975x65}

\begin{document}

\title{COSMOLOGICAL SINGULARITIES IN FLRW SPACETIMES}

\author{L. FERN\'ANDEZ-JAMBRINA}

\address{ETSI Navales, Universidad Polit\'ecnica de Madrid\\
Arco de la Victoria s/n, E-28040-Madrid, Spain\\
$^*$E-mail: leonardo.fernandez@upm.es\\
www.etsin.upm.es/ilfj.htm}

\author{R. LAZKOZ}

\address{F\'\i sica Te\'orica, Facultad de Ciencia y Tecnolog\'\i a,\\ Universidad del
Pa\'\i s Vasco, Apdo.  644, E-48080 Bilbao, Spain\\
E-mail: ruth.lazkoz@ehu.es\\
tp.lc.ehu.es/RLS.html}

\begin{abstract}
In this talk we review the appearance of new types of singularities 
(big rip, sudden singularities\ldots)
in FLRW cosmological models that have arisen on considering 
explanations for accelerated expansion of our universe.
\end{abstract}


\bodymatter

\section{Introduction}
During the last decade the observational evidence in favour of
accelerated expansion of our universe has increased considerably
\cite{accel}.  In order to cope with this fact two lines of
development have been proposed: either general relativity is to be
modified \cite{modgrav} or our universe is pervaded by an exotic
source named dark energy \cite{darkfluid}, which is responsible for
the accelerated expansion.  Violation of classical energy conditions
by this fluid has made feasible a number of new scenarios
\cite{Nojiri:2005sx} for the final fate of our universe (big
rip\cite{Caldwell:2003vq}, sudden singularities\cite{sudden},
big brake\cite{brake}, big freeze \cite{freeze}, inaccesible
singularities\cite{mcinnes}, directional singularities\cite{hidden}, 
$w$-singularities\cite{wsing}, 
in braneworld models\cite{brane}\ldots), which had not been considered
previously.  In this talk we deal with the appearance and strength of
these singular behaviours in FLRW cosmological spacetimes from the
point of view of geodesic incompleteness.

\section{Behaviour of singularities}
We choose coordinates for which the FLRW metric is written as
\[ds^2=-dt^2+a^2(t)\left\{f^2(r)dr^2+ r^2d\Omega^2\right\}, \qquad
f^2(r)=\frac{1}{1-kr^2}\;,\quad  k=0,\pm1,\]
and we assume that the scale factor $a(t)$ admits a generalised power 
expansion\cite{visser} in the vicinity of an event at $t_{0}$,
\[	a(t)=c_{0}|t-t_{0}|^{\eta_{0}}+c_{1}|t-t_{1}|^{\eta_{1}}+\cdots\;,
    \]
where $\eta_{0}<\eta_{1}<\cdots$, $c_{0}>0$. Models with oscillatory 
behaviour\cite{wands} might drop out of this scheme. 
At lowest order three different qualitative behaviours arise:
\begin{itemize}
\item $\eta_{0}>0$: a vanishing scale factor at
$t_{0}$ means a Big Bang or Big Crunch.

\item $\eta_{0}=0$: a finite scale factor at $t_{0}$ means that 
either $a(t)$ is analytical and the event is regular or on the other 
hand a weak or sudden singularity appears.

\item  $\eta_{0}<0$: a diverging scale factor at $t_{0}$ means a Big 
Rip singularity.
\end{itemize}

Geodesics are parametrised by their proper time $\tau$,
$\big(t(\tau),r(\tau),\theta(\tau), \phi(\tau)\big)$, 
\[     t'=\sqrt{\delta +\frac{P^2}{a^2(t)}}, \qquad
    r'=\pm\frac {P}{a^2(t)f(r)},\]
may be simplified in terms of a constant of geodesic motion $P$ and 
$\delta=0,1$ for null 
and timelike geodesics. The null case  $\delta=0$ is straightforwardly solved
\[ a(t) t'=P \Rightarrow \int_{t_{0}}^t
a(t')\,dt'=P(\tau-\tau_{0}),\]and the geodesic is complete if and 
only if the scale factor is an integrable function.

Dealing with finite bodies, a singularity is strong\cite{strong} if tidal forces are
capable of disrupt them.  Following Tipler, a singularity is strong if
their volume tends to zero. Folllowing Kr\'olak, just a negative
derivative of the volume is required. The results are consigned in 
Table \ref{leotable}\cite{puiseux}. Conformal diagrams for these 
models can be found in\cite{scott}.
\begin{table}
\tbl{Strength of singularities in FLRW models.}
{\begin{tabular}{cccccc}
	    \toprule
	    ${\eta_{0}}$ & ${\eta_{1}}$ & ${k}$ & $c_{0}$ &\textbf{Tipler} &
	    \textbf{Kr\'olak}  \\
	    \colrule
	    $(-\infty,0)$ & $(\eta_{0},\infty)$ &  $0,\pm 1$  & $(0,\infty)$& 
	    Strong & Strong  \\
	    $0$ & $(0,1)$ &  $0,\pm 1$  & $(0,\infty)$ & Weak & Strong  \\
	0	 & $[1,\infty)$ & $0,\pm 1$  &$(0,\infty)$& Weak & Weak
    \\
	    $(0,1)$ & $(\eta_{0},\infty)$ &  $0,\pm 1$  & 
	    $(0,\infty)$  & Strong & Strong  \\
	    $1$ & $(1,\infty)$ & $0,1$  &  $(0,\infty)$& Strong & Strong  \\
	    1 & $(1,\infty)$ & $-1$ &$(0,1)\cup(1,\infty)$ & Strong & Strong  \\
	    1 & $(1,3)$ & $-1$ & 1 & Weak & Strong  \\
	    1 & $[3,\infty)$ & $-1$ & 1 &Weak & Weak  \\
	    $(1,\infty)$ & $(\eta_{0},\infty)$ & $0,\pm 1$  &$(0,\infty)$&
    Strong & Strong \\
	    \botrule\end{tabular}}\label{leotable}\end{table}
\section{Conclusions}
 We have obtained a classification of singular events in FLRW models
 in terms of their strength and the exponents of an expansion of the scale factor around
 an event at $t_{0}$. In models with $\eta_{0}\le -1$, null geodesics avoid the
Big Rip singularity. The only models which allow regular behaviour close to  
$t_{0}$ are those with $\eta_{0}=0$ (de Sitter, sudden\ldots) and with 
$\eta_{0}=1$, $k=-1$, $c_{0}=1$ (Milne).

%
%

\section*{Acknowledgements}
R.L. is supported by the University of the Basque Country through
research grant GIU06/37 and by the Spanish Ministry of Education and
Culture through research grant FIS2007-61800.

\end{document}